\documentclass{elsart3p}

\usepackage{graphicx}

\usepackage{amssymb}

\begin{document}

\begin{frontmatter}



\title{Effect of Zn substitution for Cu on Ca$_{2-x}$Na$_{x}$CuO$_{2}$Cl$_{2}$\\ 
near the hole concentration of 1/8 per Cu}


\author[SKD]{Kohki H. Satoh,\corauthref{cor}}
\corauth[cor]{el. +81-29-879-6028,
Fax: +81-29-864-5623,
email: ksatoh@post.kek.jp}
\author[SKD]{Masatoshi Hiraishi}
\author[SKD]{Masanori Miyazaki}
\author[KEK]{Soshi Takeshita}
\author[SKD,KEK]{Akihiro Koda}
\author[SKD,KEK]{Ryosuke Kadono}
\author[KU]{Ikuya Yamada}
\author[KU]{Kengo Oka}
\author[KU]{Masaki Azuma}
\author[KU]{Yuichi Shimakawa}
\and \author[KU]{Mikio Takano}

\address[SKD]{Department of Materials Structure Science, 
The Graduate University for Advanced Studies (Sokendai), 
Tsukuba, Ibaraki 305-0801, Japan}
\address[KEK]{Institute of Materials Structure Science, 
High Energy Accelerator Research Organization (KEK), 
Tsukuba, Ibaraki 305-0801, Japan}
\address[KU]{Institute for Chemical Research, 
Kyoto University, Uji, Kyoto 611-0011, Japan}

\begin{abstract}

A weakening of superconductivity upon substitution of Cu by Zn (0.5--1\%) 
is observed in a high-$T_{\rm c}$ 
cuprate, Ca$_{2-x}$Na$_{x}$CuO$_{2}$Cl$_{2}$, near the hole concentration of 1/8 per Cu.
The superconducting transition temperature and its volume fraction, estimated by 
magnetic susceptibility, exhibit a sizable anomaly for $x=0.12$--0.14, where 
the slowing down of Cu spin fluctuations below 5~K is demonstrated by muon spin relaxation
experiments.  These observations are in close resemblance to other typical cuprates 
including YBa$_{2}$Cu$_{3}$O$_{7-\delta}$, and Bi$_{2}$Sr$_{2}$Ca$_{1-x}$Y$_{x}$Cu$_{2}$O$_{8+\delta}$, providing further evidence that Zn-induced ``stripe" correlation is a universal feature 
of high-$T_{\rm c}$ cuprate superconductors common to that of La$_{2-x}A_{x}$CuO$_{4}$ ($A$=Ba, Sr).

\end{abstract}

\end{frontmatter}

\section{Introduction}
\label{sec:introduction}

It is established that the superconducting transition temperature ($T_{\rm c}$) is suppressed 
near the hole concentration $p \sim$1/8 in La$_{2-x}$Ba$_{x}$CuO$_{4}$ (LBCO) 
and La$_{2-x-y}$Nd$_{y}$Sr$_{x}$CuO$_{4}$ (LNSCO), where
a charge- and spin-stripe order develops in place of superconductivity\cite{Tranquada,Nachumi}.
A similar tendency has been reported for La$_{2-x}$Sr$_{x}$CuO$_{4}$ (LSCO) 
with a slight shift of $x$ for the strongest strip correlation\cite{Watanabe:94}.
This so-called ``1/8 anomaly" has drawn considerable attention 
in view of a potential link to the mechanism of high-$T_{\rm c}$ superconductivity in 
cuprates\cite{Kivelson}.
It has been shown that substitution of Cu by small amounts of Zn stabilize this
stripe correlations, as it leads to the ``1/8 anomaly" in a variety of cuprates 
including LSCO\cite{Watanabe2}, YBa$_{2}$Cu$_{3}$O$_{7-\delta}$ (YBCO)\cite{Akoshima}, and Bi$_{2}$Sr$_{2}$Ca$_{1-x}$Y$_{x}$Cu$_{2}$O$_{8+\delta}$ (BSCCO)\cite{Watanabe}.
However, it is still controversial whether this anomaly is a common feature in
cuprates or that unique to the La214 system.  

Here, we report on the 1/8 anomaly in Ca$_{2-x}$Na$_{x}$-CuO$_{2}$Cl$_{2}$ (Na-CCOC)
studied by magnetic susceptibility and muon spin relaxation ($\mu$SR).
While Na-CCOC has a similar structure to La$_{2-x}$Sr$_{x}$CuO$_{4}$,
it is characterized by flat CuO$_{2}$ planes even at low temperatures owing to 
substitution of apical oxygen with chlorine in the CuO$_{6}$ octahedra.
This is in marked contrast to the case of LSCO or LBCO that have
a periodic distortion of the CuO$_{2}$ planes at low temperatures. 
The excellent cleavability of Na-CCOC crystals makes it feasible to investigate the 
electronic properties using surface-sensitive
measurements such as angle resolved photoemission spectroscopy (ARPES)
and scanning tunneling microscopy and spectroscopy (STM/STS).
The latter reports the occurrence of a checkerborad-like electronic modulation and the coexistence of 
charge order and superconductivity\cite{Hanaguri}, which exhibits a good correspondence 
with the inhomogeneous (spin glass-like) magnetic ground state revealed by $\mu$SR\cite{Ohishi}.
Meanwhile, the previous search for the 1/8 anomaly in Na-CCOC came to a negative
result\cite{Hirai}.

Considering that Zn substitution enhances stripe correlations in other cuprates, 
we prepared Zn-free and Zn substituted Na-CCOC samples 
near the hole concentration of 1/8 per Cu, and searched for the 1/8 anomaly in this compound
by means of magnetization and muon spin relaxation ($\mu$SR) measurements.

\section{Experimental details}
\label{sec:experimental}

Polycrystalline samples of Na-CCOC were prepared by high-pressure synthesis techniques.
Ca$_{2}$CuO$_{2}$Cl$_{2}$ and 
Ca$_{2}$Cu$_{0.9925}$Zn$_{0.0075}$O$_{2}$Cl$_{2}$,
prepared by solid state reaction of Ca$_{2}$CuO$_{3}$, CuO, ZnO,
CaCl$_{2}$ in N$_{2}$ flow with several grindings under ambient pressure,
were mixed with NaClO$_{4}$ and CuO, 
and then sealed in a cylindrical capsule made out of gold.
High pressure synthesis was performed 
using a cubic anvil-type high pressure apparatus operated 
under 6~GPa at a maximum temperature of 1000$^\circ$C.
Due to the extreme high hygroscopicity of Na-CCOC,
all manipulations of the samples were carried out in glove-boxes
filled with Ar.
All samples (with varying $x$) were confirmed to be of single phase 
by means of X-ray diffraction measurements.
Their structure is $I$4/$mmm$ space group 
at room temperature, where the lattice parameters change continuously as $x$ varies. 
More specifically, the $a$-axis shrinks as the Na concentration increases, indicating that 
hole carriers are introduced into the CuO$_{2}$ plane while the $c$-axis expands.
The magnetic susceptibility was measured 
using a superconducting quantum interference device
(MPMS, Quantum Design Co.), where the measurements were
made while the temperature was scanned upwards after field cooling of 20~Oe.
Conventional $\mu$SR measurements were performed on the
M15 beamline of TRIUMF (Vancouver, Canada).

\section{Result and Discussion}
\label{sec:resultsanddiscussions}

\begin{figure}[htb]
\centering
\includegraphics[width=0.87\columnwidth]{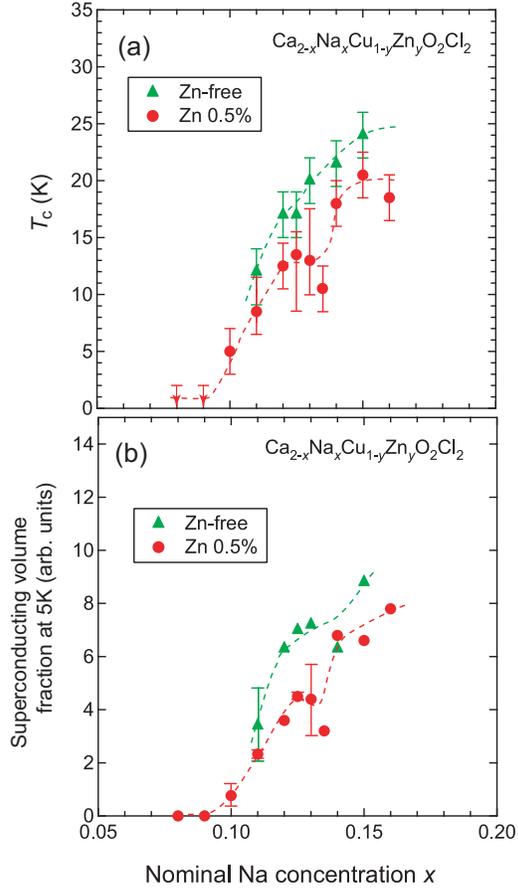}
\caption{\label{Tc} (Color online)
(a) Superconducting transition temperature ($T_{\rm c}$) versus nominal Na concentration ($x$) in 
Ca$_{2-x}$Na$_{x}$Cu$_{1-y}$Zn$_{y}$O$_{2}$Cl$_{2}$.
Triangles and circles are $T_{\rm c}$ in Zn-free samples
and in Zn 0.5~$\%$ ($y$=0.005) substituted samples, respectively.
(b) Superconducting volume fraction estimated by the magnetic
susceptibility at 5~K.
Dashed lines are giudes to eyes. }
\end{figure}

Fig.~\ref{Tc}~(a) shows the Na doping dependence of $T_{\rm c}$, where
$T_{\rm c}$ is defined as the temperature at which the 
temperature gradient of the magnetic susceptibility, $d \chi/dT$, is at its maxima. 
The error bars in Fig.~\ref{Tc}~(a) are evaluated from the spread of $\chi$
around $T_c$ and that of $T_{\rm c}$ itself among several samples with the same $x$.
For Zn-free samples, $T_{\rm c}$ exhibits a monotonous increase with Na doping
between $x$=0.11 and 0.15. 
In contrast, an overall reduction of $T_{\rm c}$ is observed for Zn
substituted samples ($y=0.005$) as compared with Zn-free ones.
Moreover, $T_{\rm c}$ shows a clear trend of leveling off around 13~K  
for 0.12 $\leq x \leq$ 0.135, which is followed by a jump to $\sim$18 K at $x$=0.14. 
This step-like behavior is similar to the one observed in Zn substituted 
YBa$_{2}$Cu$_{3}$O$_{7-\delta}$ \cite{Akoshima}.
Fig.~\ref{Tc}~(b) shows the Na doping dependence of the superconducting 
volume fraction estimated from the diamagnetic susceptibility at 5~K.
The fraction in Zn-substituted samples is also reduced as compared with 
that in Zn-free samples. 
Furthermore, a step-like change is also observed near 1/8 hole
concentration in line with the case of $T_{\rm c}$.
Thus, the 1/8 anomaly is clearly observed as anomalies of both $T_c$ and the superconducting 
volume fraction in Na-CCOC upon Zn-substitution for Cu.

We have performed $\mu$SR experiments to investigate the microscopic 
details of this anomaly.  As has been demonstrated in earlier 
reports\cite{Watanabe2,Akoshima,Watanabe},
$\mu$SR serves as a sensitive local magnetic probe that covers a unique
time window of observation ($10^{-9}$--$10^{-5}$ s) complementary to 
other magnetic probes like neutron diffraction and nuclear magnetic resonance. 
It does not rely on the long-range coherence of any magnetic order, and therefore
is useful to examine the random local magnetism, e.g., a spin-glass state.

\begin{figure}[htb]
\centering
\includegraphics[width=0.83\columnwidth]{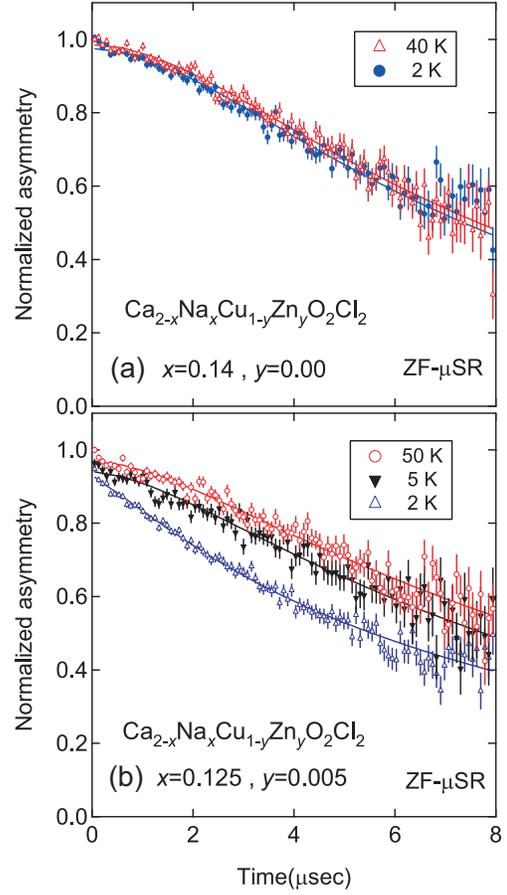}
\caption{\label{ZF} (Color online)
ZF-$\mu$SR spectra in
Ca$_{2-x}$Na$_{x}$Cu$_{1-y}$Zn$_{y}$O$_{2}$Cl$_{2}$ 
with (a) $x$=0.14,$y$=0.00 
and (b) $x$=0.125, $y$=0.005.
Solid curves represent fits to the data using equation (1).}
\end{figure}

Fig.~\ref{ZF} shows $\mu$SR spectra measured under zero external 
field conditions (ZF) in Ca$_{2-x}$Na$_{x}$Cu$_{1-y}$Zn$_{y}$O$_{2}$Cl$_{2}$ 
for  (a) Zn-free samples ($x=0.14$, $y=0.00$) and (b) 0.5\% of Zn substitution for Cu
($x=0.125$, $y=0.005$).  We observe a Gaussian depolarization due to random 
local fields from nuclear moments in the Zn-free
samples over the entire temperature range down to 2~K.
Meanwhile, in the latter case, an exponential damping is observed in the compound 
with Zn-substitution at 2~K. 
The spectra suggest a slowing down of Cu spin fluctuations with decreasing
temperature below $\sim$5 K.
A similar phenomenon associated with Zn substitution was also reported for the case of
the La214 systems, YBCO, and BSCCO,
and it suggests that the Zn impurity effect is a common feature 
of High-$T_{\rm c}$ cuprate superconductors\cite{Watanabe2,Akoshima,Watanabe}. 
However, it must be noted that the Cu spins are not completely static nor in any long-range ordered 
state at 2 K, as inferred from 
the absence of a spontaneous oscillatory signal in the $\mu$SR time spectrum.
This might be because the temperature is still too high to freeze out the Cu spins.

The $\mu$SR time spectra (=asymmetry) are analyzed by fits using the following form,
\begin{equation}
AP(t) = [A_{1}+A_{2}{\rm exp}(-\lambda t)]G_{\rm DKT}(\Delta, \nu, t), 
\end{equation}
where the first term represents a signal from muons stopping in a non-magnetic region.
The second term represents that of magnetic regions in which the muon senses 
Cu spin fluctuations, $A_{i}$ are the partial asymmetries which are proportional to 
the respective volume fractions,  $\lambda$ is the depolarization rate, and 
$G_{\rm DKT}(\Delta, \nu, t)$ is the Kubo-Toyabe function that represents 
the Gaussian damping due to nuclear random local fields
(with $\Delta$ being the linewidth and $\nu$ the fluctuation rate
of the nuclear random local fields). 

\begin{figure}[tb]
\centering
\includegraphics[width=0.9\columnwidth]{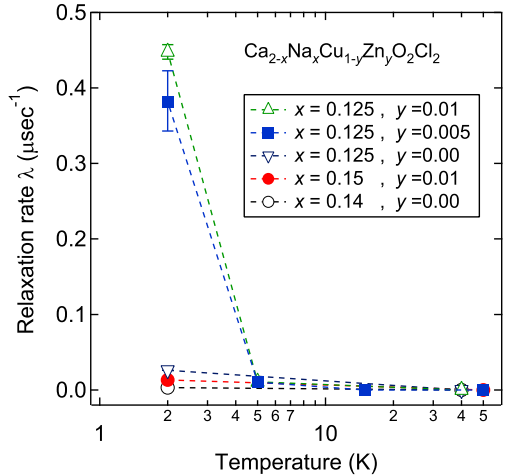}
\caption{\label{rlx} (Color online)
Temperature dependence of the muon spin depolarization rate ($\lambda$) in 
Ca$_{2-x}$Na$_{x}$Cu$_{1-y}$Zn$_{y}$O$_{2}$Cl$_{2}$ with a
variety of Na ($x$) and Zn ($y$) concentrations.}
\end{figure}

As shown in Fig~\ref{rlx}, $\lambda$ is almost zero in Zn-free samples 
[$x$=0.14, $y$=0.00 ($T_{\rm c}\sim$ 22~K),
$x$=0.0125, $y$=0.00 ($T_{\rm c}\sim$ 17~K)]
and in the sample with $x \neq 1/8$
[$x$=0.15, $y$=0.01($T_{\rm c}\sim 10$ K)], where the 
non-magnetic region dominates over the entire sample volume. 
On the other hand,
it increases below 5~K in Zn-substituted samples near $x \sim$1/8
[$x$=0.0125, $y$=0.005 ($T_{\rm c}\sim$ 13~K) and
$x$=0.0125, $y$=0.01 (not superconducting)]
Thus, it is inferred from ZF-$\mu$SR experiments that the 1/8 anomaly observed in 
bulk properties is associated with the slowing down of Cu spin 
fluctuations  in Na-CCOC.

In LBCO and LNSCO near $x\simeq1/8$, the appearance of an incommensurate spin 
density wave (SDW) phase and an associated suppression of superconductivity has been 
reported within the present range of temperature
\cite{Tranquada,Tranquada2}.
The absence of (or the reduction of the characteristic temperature for) such a static SDW 
phase in Na-CCOC may be attributed to that of the structural phase transition which is
known to occur in La214 system\cite{Katano}.   In LBCO at $x$=1/8, the lattice 
structure exhibits a successive change with decreasing temperature from 
a high-temperature tetragonal (HTT) phase at room temperature 
to a low-temperature orthorhombic (LTO) phase, and then to a 
low-temperature tetragonal (LTT) phase.
Each phase is different in symmetry with respect to the distortion along the 
CuO$_{2}$ planes, and the SDW phase occurs in accordance with the LTT phase in LBCO.  
Meanwhile, in the case of LSCO that does not exhibit the LTT phase,
the suppression of superconductivity is relatively weak.
Since Na-CCOC remains in the HTT phase over the entire temperature 
range of observation ($>2$ K),  the absence of the static SDW phase near $x=1/8$
may suggest a common trend that the LTT phase stabilizes the SDW phase
(by ``pinning the dynamical stripe").
However, it is repoted that Zn substitution for Cu enhances the SDW state in LSCO\cite{Watanabe2},
and that the SDW phase survives in spite of the suppression of the LTT phase by applying
hydrostatic pressure\cite{Satoh}.
These observations might in turn point to the reversed role of 
cause and effect between the instability towards the SDW phase and the occurrence of LTT phase. 
The present result would make a strong case for the SDW instability as 
an intrinsic feature of the CuO$_2$ planes irrespective of the LTT phase.

It might be speculated that the electronic state of the Cu ions are sensitive 
to the {\sl local} lattice distortion (but not to the long range lattice morphology) 
so that the local impurities like Zn have relatively strong influence on the 
SDW instability. In this regard, another factor would be the degree of A-site disorder 
that is known to be different between LBCO and LSCO. 
Eisaki $et~al.$ reported that the $A$-site disorder has a certain influence 
on $T_{\rm c}$\cite{Eisaki}, suggesting that the CuO$_2$ planes are
strongly affected by the  $A$-site ions.

In conclusion, we demonstrated the presence of the 1/8 
anomaly in Ca$_{2-x}$Na$_{x}$CuO$_{2}$Cl$_{2}$ with  
a small fraction of Cu substituted by Zn.
Such an anomaly is also observed in many families of cuprates
near the hole doping concentration of 1/8,
and it suggests that the SDW instability against local distortion of 
the CuO$_2$ planes is a common feature of high-$T_{\rm c}$ cuprate superconductors.

{\ }\\
{\bf Acknoledgemnts}

We would like to thank the staff of TRIUMF for their technical support during the
$\mu$SR experiment.  This work was supported by the KEK-MSL Inter-University 
Program for Oversea Muon Facilities and a Grant-in-Aid for Scientific Research
on Priority Areas by Ministry of Education, Culture, Sports, Science and Technology, Japan.


\begin{thebibliography}{99}

\bibitem{Tranquada}
J.~M.~Tranquada~$et~al$., Nature $\bf{375}$ (1995) 561.

\bibitem{Nachumi}
B.~Nachumi~$et~al$., Phys.~Rev.~B $\bf{58}$~(1998)~8760.

\bibitem{Watanabe:94}
I.~Watanabe~$et~al$., Hyperfine Interact. $\bf{86}$~(1994)~603.

\bibitem{Kivelson} 
S. A. Kivelson ~$et~al$., Rev.~Mod.~Phys.~ $\bf{75}$~(2003)~1201.

\bibitem{Watanabe2}
I.~Watanabe~$et~al$., J.~Phys.~Chem.~Solids $\bf{63}$~(2002)~1093.

\bibitem{Akoshima}
M.~Akoshima~$et~al$., Phys.~Rev.~B $\bf{62}$~(2000)~6761.

\bibitem{Watanabe}
I.~Watanabe~$et~al$., Phys.~Rev.~B $\bf{62}$~(2000)~14524.

\bibitem{Hanaguri}
T.~Hanaguri~$et~al$., Nature $\bf{430}$ (2004) 1001.

\bibitem{Ohishi}
K.~Ohishi~$et~al$., J.~Phys.~Soc.~Jpn., $\bf{74}$~(2005)~2408.

\bibitem{Hirai}
D.~Hirai~$et~al$., Physica C $\bf{463-465}$~(2007)~56.

\bibitem{Tranquada2}
J.~M.~Tranquada~$et~al$., Phys.~Rev.~B $\bf{54}$~(1996)~7489.

\bibitem{Katano}
S.~Katano~$et~al$., Phys.~Rev.~B $\bf{48}$~(1993)~6569.

\bibitem{Satoh}
K.~H.~Satoh~$et~al$., Physica B $\bf{374-375}$~(2006)~40.

\bibitem{Eisaki}
H.~Eisaki~$et~al$., Phys.~Rev.~B $\bf{69}$~(2004)~064512.



\end{thebibliography}
\end{document}